\documentclass[a4paper,12pt,article,aps]{revtex4}
\usepackage{amssymb,amsmath}
\usepackage{textcomp}
\usepackage{indentfirst}
\usepackage[mathscr]{eucal}
\usepackage{graphicx} %

\begin{document}

\title{Vibrational resonances in 1D Morse and FPU lattices.}

\author{T. Yu.~Astakhova} %
\email{astakhova@deom.chph.ras.ru} %
\affiliation{N. M.~Emanuel  Institute of Biochemical Physics, \\
Russian Academy of Sciences, ul.~Kosygina~4, Moscow 119334,
Russia}

\author{N. S.~Erikhman} %
\email{erikhman@deom.chph.ras.ru} %
\affiliation{N. M.~Emanuel  Institute of Biochemical Physics, \\
Russian Academy of Sciences, ul.~Kosygina~4, Moscow 119334,
Russia}

\author{V. N.~Likhachev} %
\email{vl@deom.chph.ras.ru} %
\affiliation{N. M.~Emanuel  Institute of Biochemical Physics, \\
Russian Academy of Sciences, ul.~Kosygina~4, Moscow 119334, Russia} %

\author{G. A.~Vinogradov
 {\footnote {To whom all correspondence should be addressed.}}} %
\email{gvin@deom.chph.ras.ru. } %
\affiliation{N. M.~Emanuel  Institute of Biochemical Physics, \\
Russian Academy of Sciences, ul.~Kosygina~4, Moscow 119334, Russia} %


\renewcommand\baselinestretch{1.25}\normalsize

\begin{abstract}


In the present paper the resonances of vibrational modes in
one-dimensional random Morse lattice are found and analyzed. The
resonance energy exchange is observed at some values of
elongation. Resonance $2 \, \omega_1 = \omega_2$ is investigated
in details. The interacting modes are inequivalent: the
higher-frequency mode is much more stable in the excited state,
i.e. its life-time is larger than the life-time of
lower-frequency mode under the resonance conditions. Simple model
of two nonlinearly coupled harmonic oscillators is also
considered. It allows to get analytical description and to
investigate the kinetics and the energy exchange degree vs. such
parameters as the resonance detuning and specific energy. The
very similar behavior is found in the Morse and the
two-oscillatory models, and an excellent agreement between
analytical and numerical results is obtained. Analogous resonance
phenomena are also found in the random Fermi-Pasta-Ulam lattice
under contraction.

\vspace{0.5 cm}





\end{abstract}

\maketitle

\section{Introduction }%
  \label{sec:Intro}

The primary goal of Fermi, Pasta, Ulam and {\emph{Tsingou}}
(FPUT) \cite{FPU55}
\footnote{Th.~Dauxois highly recommends \cite{Dau08} to quote
from now on the Fermi-Pasta-Ulam-{\emph {Tsingou}} problem taking
into account the great contribution of Mary Tsingou in the
formulation and solving the FPUT--problem.}
was to find an energy sharing in one-dimensional lattices with
nonlinear interaction between neighboring particles. The authors
expected the occurrence of the statistical behavior in a system
with coupled nonlinear oscillators due to the nonlinear
interaction, leading to the energy equidistribution among the
degrees of freedom. Initially, the long-wave vibrations were
excited and {\emph {``Instead of a gradual increase of all the
higher modes, the energy is exchanged, essentially, among only a
certain few. It is, therefore, very hard to observe the rate of
\,"thermalization"\, or mixing in our problem, and this was the
initial purpose of the calculation.''}} \cite{FPU55}.

Much effort has been undertaken to explain the FPUT results. Two
approaches were developed. The first one was to analyze the
dynamics of the nonlinear lattice in the continuum limit, which
led to the discovery of solitary waves \cite{Zab65}. The second
approach, advanced by Izrailev and Chirikov, pointed to the
overlap of nonlinear resonances \cite{Chi60} and an existence of
a stochasticity threshold in the FPUT system \cite{Izr66}. For
strong nonlinearities (or/and large energies) the overlap of
nonlinear resonances results in the dynamical chaos which
destroys the FPUT recurrence and ensures convergence to thermal
equilibrium. (For more details on the FPUT problem see recent
reviews devoted to the 50th anniversary of the celebrated paper
\cite{FPU55} in {\it Focus Issue: The Fermi-Pasta-Ulam Problem
--- The First Fifty Years}, ed. by D.K. Campbell, P. Rosenau, and
G.M. Zaslavsky, Chaos {\bf 15}(1) (2005)\,).

In the present paper we thoroughly analyze the resonance between
the lowest $(k=1)$ and the next $(k=2)$ vibrational modes
previously discovered in the random Morse lattice under the
elongation \cite{Ast07}. Here we also employ the lattice with the
random interaction potential between neighboring particles under
different deformation degrees. The lattice randomness makes the
vibrational frequencies to be irregular, and the deformation
allows to achieve the true resonance conditions. Random FPUT
lattice (with random quadratic terms) \cite{Liv85} and with
alternative masses \cite{Zab67} were investigated but no
resonances were directly observed.

The model of two nonlinearly coupled harmonic oscillators with
the frequencies ratio $\omega_1 \! : \! \omega_2 = 1 \! : \! 2$
is also considered. An excellent agreement in dynamics and
kinetics of the Morse lattice and the oscillatory model is
observed.

The random $\alpha$--FPUT lattice under compression demonstrates
the similar behavior as the Morse lattice. The compression of the
$\alpha$--FPUT lattice in contrast to the elongation of the Morse
lattice is explained by the different signs of cubic terms in an
expansion of the Morse potential $(\widetilde U_{\text{M}} = x^2
- x^3)$ and in the $\alpha$--FPUT lattice $U = x^2/2 + \alpha
x^3/3$.


\section{Resonances in random Morse lattice and two-oscillatory model.}
  \label{sec:Morse_and_2}

\subsection{Random Morse lattice.}
  \label{subsec:Morse}

\vspace{-0.3 cm}

We consider the finite--length lattice of $N$ particles
interacting via the Morse potential with the rigid boundaries. The
potential energy has the form:
\begin{equation}
\label{U_p} %
U_{\text p} = \sum_{i=1}^{N+1} \xi_i\,U(y_i); \qquad y_i = x_i -
x_{i-1} \quad (x_0 = 0, \,\, x_{N+1} = L),
\end{equation}
where $x_i$ is the displacement of $i$th particle from the
equilibrium and $U(y) = \left[ 1- \exp(-y) \right]^2$ is the
dimensionless Morse potential. $\xi_i$ is a random number chosen
from the range $[0{.}5 - 1{.}5]$ and is modeling the random well
depth of the interaction potential. The left lattice end is fixed
and the right end has an arbitrary coordinate $L > 0$. The
specific lattice elongation is $\varepsilon = L/(N+1)$.

\begin{figure*}
\begin{center}
\includegraphics[width=100mm,angle=0]{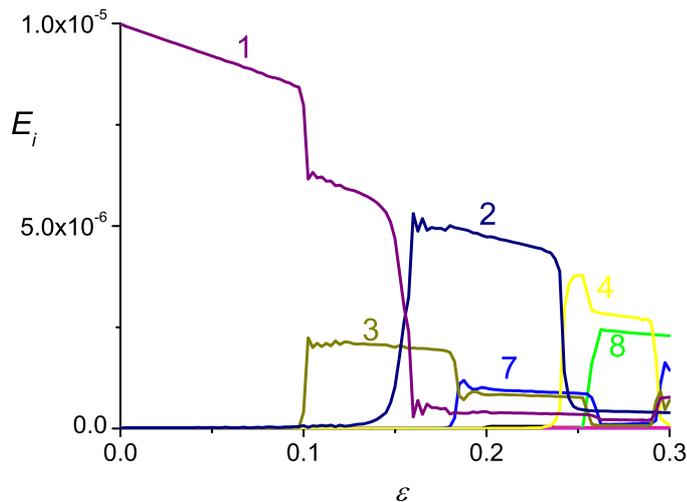}
\end{center}
\vspace{-1.5 cm} \caption{ \label{fig_1} Evolution of the
vibrational modes energies vs. the specific elongation
$\varepsilon$. The most long-wave mode ($k=1$) is initially
excited with the energy $E_0=10^{-5}$. Digits numerate the modes
numbers. $N=10$. Modes not shown ($ k =5,6,9,10$) are not
excited.}
\end{figure*}

The true resonance condition ($\sum_k m_k \omega_k = 0$; $\{ m \}
\in \mathbb Z$ \cite{For61}), can be achieved at some elongation
values (apparently dependent on the set $\left\{ \xi_i \right\}$
of random values in \eqref{U_p}) as it is demonstrated in
Fig.~\ref{fig_1}. The most long-wave mode ($k=1$) with the energy
$E_0=10^{-5}$ was initially excited and the energies of
vibrational modes under adiabatically slow elongation are shown.
Small number of particles $N=10$ and low excitation energy allow
to avoid the resonances overlap.
One can see that the energy exchange is observed at some values
of $\varepsilon$. For instance, pair resonances $\omega_3 \! : \!
\omega_1 = 3$ ($\varepsilon \approx 0.101$), $\omega_2 \! : \!
\omega_1 = 2$ ($\varepsilon \approx 0.154$) are obvious.
Moreover, triple resonances are also presented (see e.g.
$\omega_3 + \omega_4 = \omega_8$ at $\varepsilon \approx 0.255$).
The energy exchange rates and degrees depend on $E_0$ and the
elongation rate. The energy decrease in Fig.~\ref{fig_1} is
explained by the ``softening'' of the potential $U$ in
\eqref{U_p} under the elongation \cite{Lik06}, i.e. under
adiabatic deformations the mode energy $E_i$ decreases as its
frequency: $E_i \sim \omega_i$, and $w \sim \sqrt{g}$, where $g$
-- lattice rigidity.

The resonance $\omega_2 \! : \! \omega_1 = 2$ at $\varepsilon
\approx 0.154$ is chosen for detailed investigation. But firstly
we consider the two-oscillatory model, where the resonance
behavior can be analytically described.


\subsection{Two-oscillatory model.}
  \label{subsec:two-osc}

\vspace{-0.3 cm}

We consider two nonlinearly coupled harmonic oscillators with the
following hamiltonian (here the main results are briefly
demonstrated, whereas the full analysis will be presented
elsewhere):
\begin{equation}
  \label{ham-a} %
  H = \dfrac12  \left(  p_1^2 +p_2^2 \right) +
      \dfrac12  \left(  \kappa_1 \, q_1^2  + \kappa_2 \, q_2^2 \right) +
      \lambda \, q_1^2 \, q_2\, ,
\end{equation}
where $\kappa_1=1$, $\kappa_2=4$, $\lambda$ is the coupling
parameter, and the frequencies ratio is 1:2.

After the variables replacements $  q_i \mapsto - \sqrt{2 \, I_i/
\omega_i} \, \cos \phi_i $ and  $(2I_1 - I_2)/5 \mapsto I, \
(2I_1 - I_2)/5 \mapsto J, \ (2 \phi_1 - \phi_2) \mapsto \Theta, \
(\phi_1 + 2\phi_2) \mapsto \Phi$ and the averaging
\cite{Bog61,Arn74} over the period $2 \, \pi/\omega_2$
Eq.~\eqref{ham-a} transforms to
\begin{equation}
  \label{ham-d} %
  {\widetilde H}(I,J) = 5 \,I - \gamma \, (I + 2 \, J) \, \sqrt{2 \,I -J \,}
  \, \cos \Theta \,,
\end{equation}
where $\gamma = \kappa_1 \kappa_2 \, \lambda$. $I$ is the integral
of motion and the time evolution is determined by the second term.
The energy exchange is governed by the following hamiltonian
\begin{equation}
  \label{ham-e} %
  \mathcal{H} = - \dfrac{{\widetilde  H} - 5 \, I}{\gamma \, I} =
  (1 + 2 \, p) \, \sqrt{2 - p} \, \cos \Theta \,,
\end{equation}
where $p = J/I$. The initial conditions $p_0=2$ and $p_0=-1/2$
correspond to the excitation of the oscillators $\omega_1=1$ and
$\omega_2=2$, correspondingly. The Hamiltonian equations for
$\mathcal{H}(p,\Theta )$ have the form:
\begin{equation}
  \label{eq-1} %
        \begin{array}{lll}
    \dot p      & = & (1+2\,p) \, \sqrt{2-p} \, \sin \Theta \,, \vspace{0.3 cm}\\
    \dot \Theta & = & \dfrac{7 - 6\,p}{2 \,\sqrt{2-p}} \, \cos
    \Theta \,.
    \end{array}
\end{equation}

\noindent {\bf 1. Excitation of the ``stable'' oscillator.} The
initial conditions $p(0) = -1/2, \ \ \Theta(0) = 0$ correspond to
the excitation of the oscillator $\omega_2 = 2$. In this case the
solution of Eq.~\eqref{eq-1} is
\begin{equation}
  \label{sol-1} %
    \begin{array}{rll}
     p(t) & = & -1/2, \\
     \sin \Theta(t) & = &
     \dfrac{\exp(2 \sqrt{10}t) - 1}{\exp(2 \sqrt{10}t) + 1}
    \end{array}
\end{equation}
and $p(t) =$ const, what means that the oscillator $\omega_2$ has
{\emph{ the infinite life-time}} and the oscillator $\omega_1$
{\emph{ is not excited at all}}.

\noindent {\bf 2. Excitation of the ``unstable'' oscillator.} The
situation drastically changes if the oscillator $\omega_1=1$ is
initially excited. Since $(p=2, \ \Theta = 0)$ is a singular
point of \eqref{eq-1}, the initial condition $p(t=0) = p_0 = 2 -
\delta$ $(\delta \ll 1)$ should be chosen with the final limiting
transition $\delta \to +0$. Taking into account that \eqref{eq-1}
has the integral of motion $\cos \Theta = \dfrac{(1 + 2 \, p_0)
\, \sqrt{2 - p_0}} {(1 + 2 \, p) \, \sqrt{2-p}}$ one can get the
following solution in the vicinity of $p(t) \sim p_0$:
\begin{equation}
  \label{sol-3}
   p(t) = p_0 - \dfrac{(12 \, p_0^2 - 8 \, p_0 - 7)}{4} \, t^2,
\end{equation}
and $p(t)$ varies through a finite range in a finite time $\sim
\sqrt{4\, \Delta p/a(p_0)}$ and the phase trajectory goes away
from the singular point $p_0 = 2$ over a finite distance. Then,
according to \eqref{eq-1}, $p(t)$ decreases approaching the value
$p = -1/2$. It means that the initially excited oscillator
$\omega_1$ loses its energy in a finite time and completely gives
the energy up to the higher frequency oscillator. This allows to
specify {\emph{the oscillator $\omega_1$ as an unstable}}, and
{\emph{the oscillator $\omega_2$ as a stable one}}.

\noindent {\bf 3. ``Mixed'' initial conditions.} Below we briefly
consider the ``mixed'' initial condition $(p_0 = -1/2 + \nu,0)$
when the stable oscillator $\omega_2$ is initially excited with
small ($\nu \ll 1$) ``weight'' of the low-frequency oscillator.
In this case the energy exchange is also observed and the
life-times of both oscillators in excited states can be estimated
as:
\begin{equation}
 \label{t-life}
 \begin{array}{l}
  \tau _1   \xrightarrow[p_0 \to  -1/2]{} \dfrac1{\sqrt{10}} \,
  \ln \left( \dfrac{\sqrt{3} + 1}{\sqrt{3} - 1}  \right); \\
  \tau _2 \xrightarrow[p_0 \to  -1/2]{} \dfrac1{\sqrt{10}} \, \ln
  \dfrac1{1+2p_0}
  \end{array}
\end{equation}
and it follows that if the stable oscillator $\omega_2$ is
initially excited with a small addition $\nu \ll 1$ of the
oscillator $\omega_1$, then the energy exchange is observed and
{\emph{the life-time of the unstable lower-frequency oscillator
$\tau_1$ is finite}} and does not depend on $\nu$, whereas
{\emph{the life-time $\tau_2$ of the stable higher-frequency
oscillator increases $\sim -\ln(\nu)$ as $\nu \rightarrow +0$}}.

\noindent {\bf 4. Resonance detuning.} Obviously, the considered
exact resonance $\omega_2 = 2 \, \omega_1$ rarely occurs and the
situation of resonance detuning is of particular interest. Let
$\kappa_1 = (1-\mu)$ and $\kappa_2 = 4$ in Eq.~\eqref{ham-a},
where $\mu \ll 1$ means a small resonance detuning. The case of
the initial excitation of the stable oscillator is trivial: the
equations of motion of \eqref{ham-a} are:
\begin{equation}
  \label{a}
 \ddot q_1 = -(1 - \mu) \, q_1 - 2 \lambda \, q_1 q_2 ; \qquad
 \ddot q_2 = - 4 \, q_2 -  \lambda \, q_1^2
\end{equation}
and if $q_1(t=0)=0, \ q_2(t=0)\neq 0$ then the first oscillator
is not excited at all, and the initially excited second
oscillator lives infinitely long independently on the vicinity to
the exact resonance.

The case of the initial excitation of the unstable oscillator
$\omega_1$  requires more accurate analysis. It is convenient to
make the variables replacements: $(q_1 \cos \alpha - q_2 \sin
\alpha) \mapsto y_1$ and $(q_1 \sin \alpha + q_2 \cos \alpha)
\mapsto y_2$. Then by neglecting the terms of the order of
$\lambda \mu \ll 1$ one can choose $\alpha$ such that the true
resonance problem is fulfilled in new variables $\{y_1,y_2\}$.
But this problem was considered above. And finally the life-times
of both oscillators are estimated as:
\begin{equation}
 \label{Tf}
 \begin{array}{l}
  \tau_1 (\mu) =  \dfrac1{\sqrt{10}} \,
  \ln \left( \dfrac{\sqrt{3}+1}{\sqrt{3}-1} \right); \\
    \tau_2 (\mu) = \dfrac1{2\sqrt{10}} \,
  \ln \left( \dfrac{3}{80 \, \mu} \right).
  \end{array}
\end{equation}
Thus if the unstable oscillator $\omega_1$ is initially excited,
then its life-time is {\emph{a finite value}} and doesn't depend
on the closeness to the exact resonance point, while the
life-time of the stable oscillator {\emph{increases}} as $\sim
-\ln(\mu)$ at $\mu \to +0$.

Below we numerically compare the kinetics and the energy exchange
degree for the Morse lattice and the two-oscillatory model.


\section{Numerical simulations of resonance interaction
in the Morse lattice and two-oscillatory model.}
 \label{sec:num}

\subsection{Comparison of the Morse lattice and two-oscillators model.}

\vspace{-0.3 cm}

Our primary goal is to investigate the rate and the energy
exchange degree near the resonance for two most long-wave modes
$(k=1,2)$ in the Morse lattice depending on the model parameters
(resonance detuning,  excitation energy). The comparison is made
with  the two-oscillatory model and one-to-one correspondence
between vibrational modes $k = 1,2$ in the Morse lattice and
lower and higher frequency oscillators is assumed.

Note, that because of the nonlinearity of both models, the
mode/oscillator frequencies depend on the excitation energy
$E_0$. And the true resonance conditions are fulfilled for some
values $\varepsilon_0(E_0)$ and $\mu_0(E_0) \neq 0$. It is
convenient to represent the actual values of $\varepsilon$ and
$\mu$ as $\varepsilon = \varepsilon_0(E_0) + \varepsilon_1$ and
$\mu = \mu_0(E_0) + \mu_1$, where $\varepsilon_1$ and $\mu_1$ are
the measures of resonances detuning.

We compare the resonance $2 \, \omega_1 = \omega_2$ at
$\varepsilon \simeq 0{.}154$ (see Fig.~\ref{fig_1}) in the Morse
lattice and the resonance in the two-oscillatory model with
$\kappa_1=1-\mu$, $\kappa_2=4$; parameter $\lambda$ for
definiteness is chosen $\lambda=0.3$. The exact resonance values
$\mu_0$ and $\varepsilon_0$ are determined as values, where the
largest life-times of the stable mode/oscillator are achieved in
numerical simulation. These values are $\varepsilon_0(E_0) =
0.153\,808\,281\,3623$ and $\mu_0(E_0) = 3{.}374\,961\,52 \cdot
10^{-8}$, and $E_0=10^{-6}$ in both cases. All digits in
$\varepsilon_0(E_0)$ are accurate: variation of the last (13{\sl
th}) digit noticeably diminishes the vibrational life-time.

Thus one can analyze and compare the interaction of vibrational
modes and oscillators vs. detuning parameters $\varepsilon_1$ and
$\mu_1$ in both systems.

\subsection{Numerical results for the Morse lattice and oscillatory
model.}

\vspace{-0.3 cm}

\begin{figure}
\begin{center}
\includegraphics[height=48mm,angle=0]{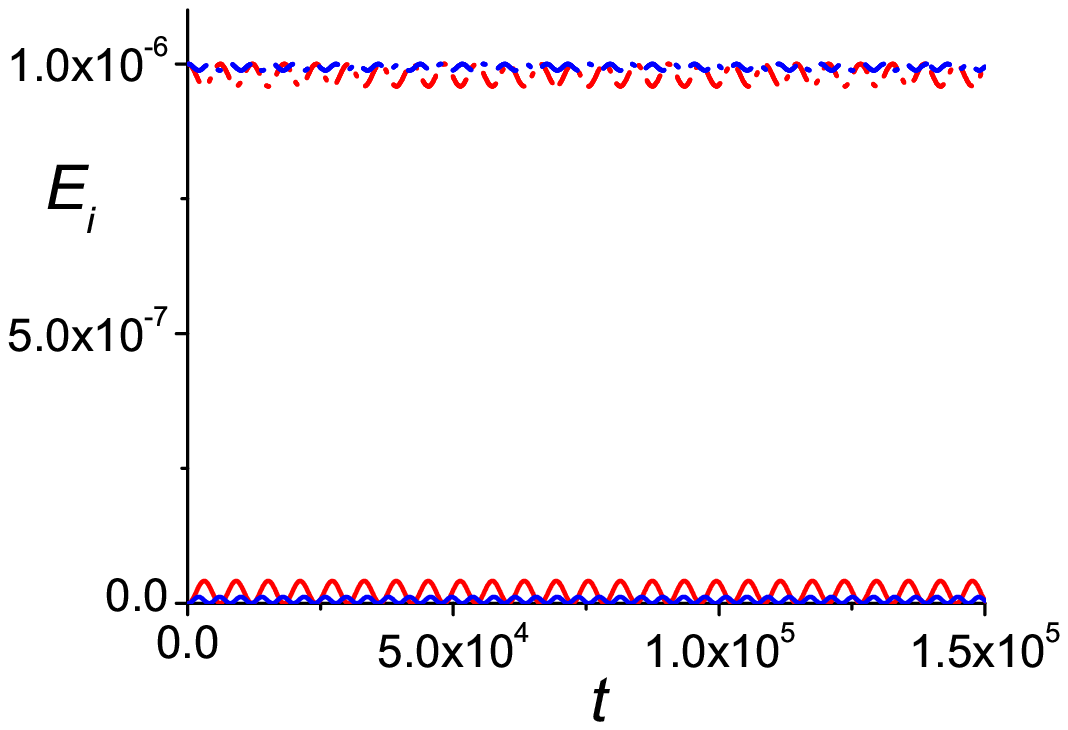}
\includegraphics[height=48mm,angle=0]{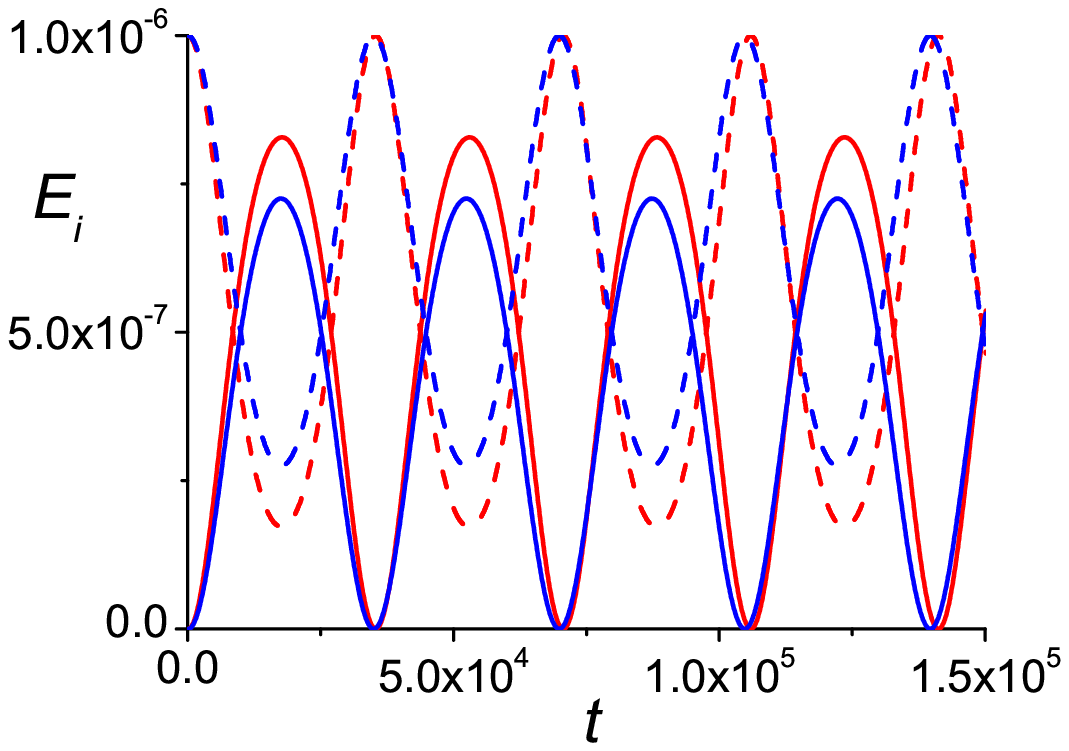}

\vspace{-0.6 cm}

a \hspace{6.5 cm} b

\includegraphics[height=48mm,angle=0]{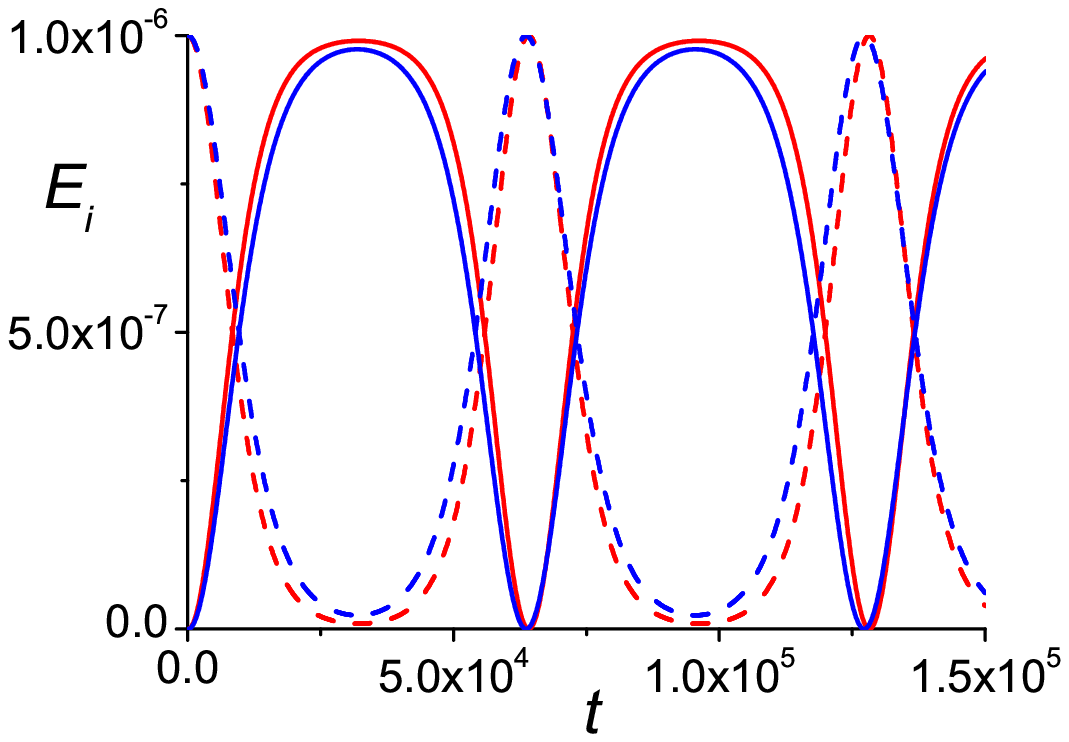}
\includegraphics[height=48mm,angle=0]{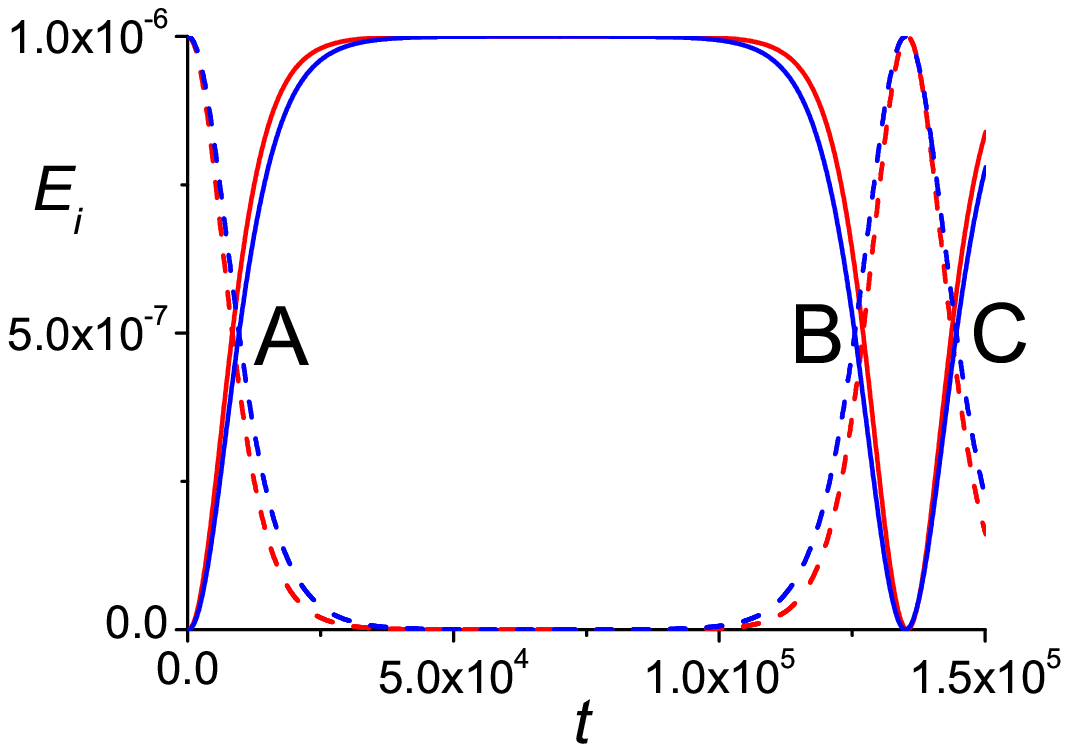}

\vspace{-0.6 cm}

c \hspace{6.5 cm} d

\includegraphics[height=48mm,angle=0]{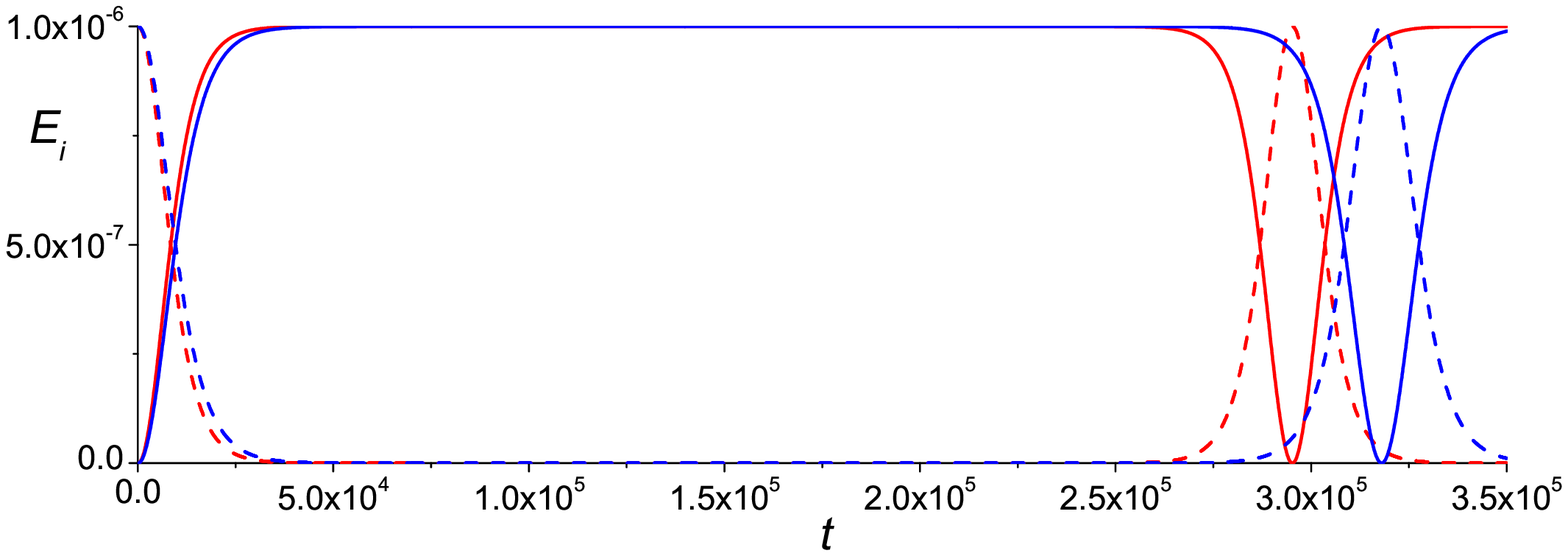}

\vspace{-0.6 cm}

e

\end{center}
\caption{
 \label{fig_2}%
Temporal modes/oscillator energies evolution at different values
of the resonance detunings for the Morse lattice (blue color) and
the two oscillators model (red color). Stable mode/oscillator --
solid lines, unstable mode/oscillator -- dashed lines.
a) $\varepsilon_1 = -2{.}380\,828\,136\,23 \cdot 10^{-2}$,
$\mu_1= 10^{-3}$;
b) $\varepsilon_1 = -8{.}082\,813\,623 \cdot 10^{-4}$, $\mu_1=4
\cdot 10^{-5}$;
c) $\varepsilon_1 =-5{.}828\,136\,23 \cdot 10^{-5}$, $\mu_1=1{.}9
\cdot 10^{-6}$; %
d) $\varepsilon_1 = -8{.}136\,23 \cdot 10^{-8}$, $\mu_1=10^{-9}$; %
e) $\varepsilon_1 = 0{.}0$, $\mu_1=0{.}0$. In all cases the
unstable mode/oscillator are initially excited with
$E_0=10^{-6}$. Note the enlarged time scale in panel e).
 }
\end{figure}

The temporal behavior of  modes/oscillators energies is shown in
Fig.~\ref{fig_2} at different values of detuning parameters
$\varepsilon_1$ and $\mu_1$. Here the unstable mode/oscillator
are initially excited with $E_0=10^{-6}$.
The general scenario of the kinetics and the energy exchange
degree is very similar for both systems.

There is no interaction between the vibrational modes/oscillators
far away from the resonance conditions ($|\varepsilon_1| \gtrsim
10^{-2}$ and $|\mu_1| \gtrsim 10^{-3}$). There exists partial and
symmetrical energy exchange in the middle range of the detuning
parameters $10^{-4} \lesssim |\varepsilon_1| \lesssim 10^{-2}$
and $10^{-6} \lesssim |\mu_1| \lesssim 10^{-3}$
(Figs.~\ref{fig_2}a-b).  As the resonance point is approached (by
diminishing $|\varepsilon_1|$ and $|\mu_1|$) the energy exchange
degree and the modes asymmetry increases. And at a certain
detuning parameters values  the full energy exchange is achieved
(Fig.~\ref{fig_2}c). Further approaching to the resonance point
results in the increase of the stable mode/oscillator life-times,
whereas the life-times of the unstable mode/oscillator do not
change (Figs.~\ref{fig_2}d-e). And finally the life-times of the
stable mode/oscillator achieve their maximal values at the
resonance point $|\varepsilon_1| = |\mu_1| =0{.}0$
(Fig.~\ref{fig_2}e).

\begin{figure*}
\begin{center}
\includegraphics[height=60mm,angle=0]{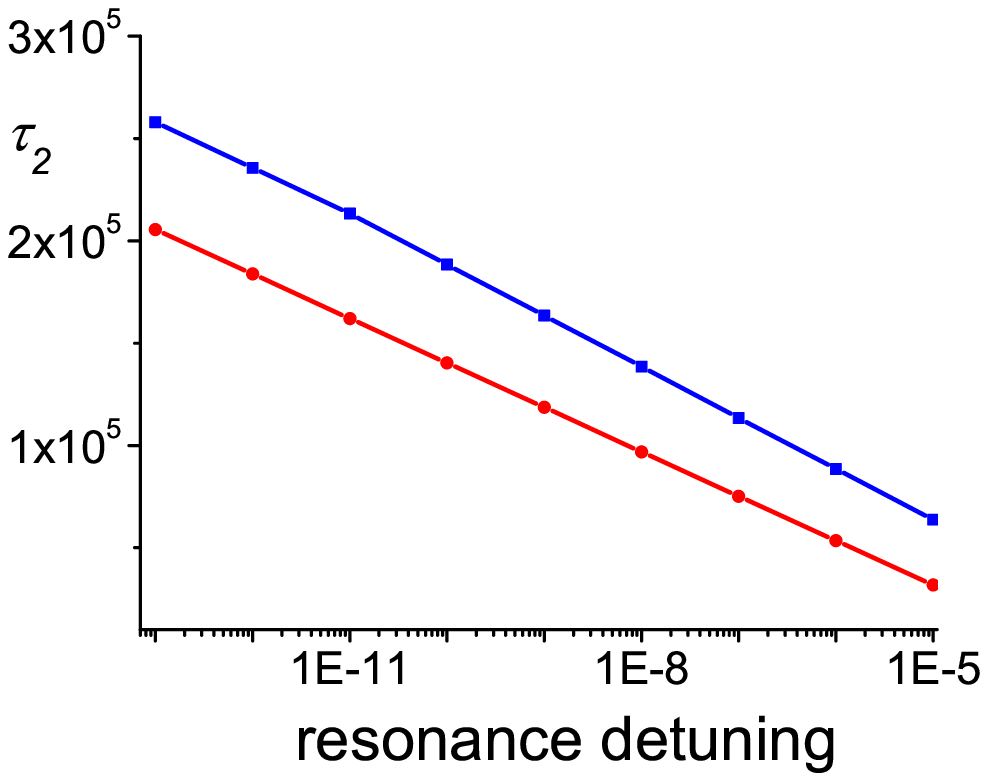}
\includegraphics[height=60mm,angle=0]{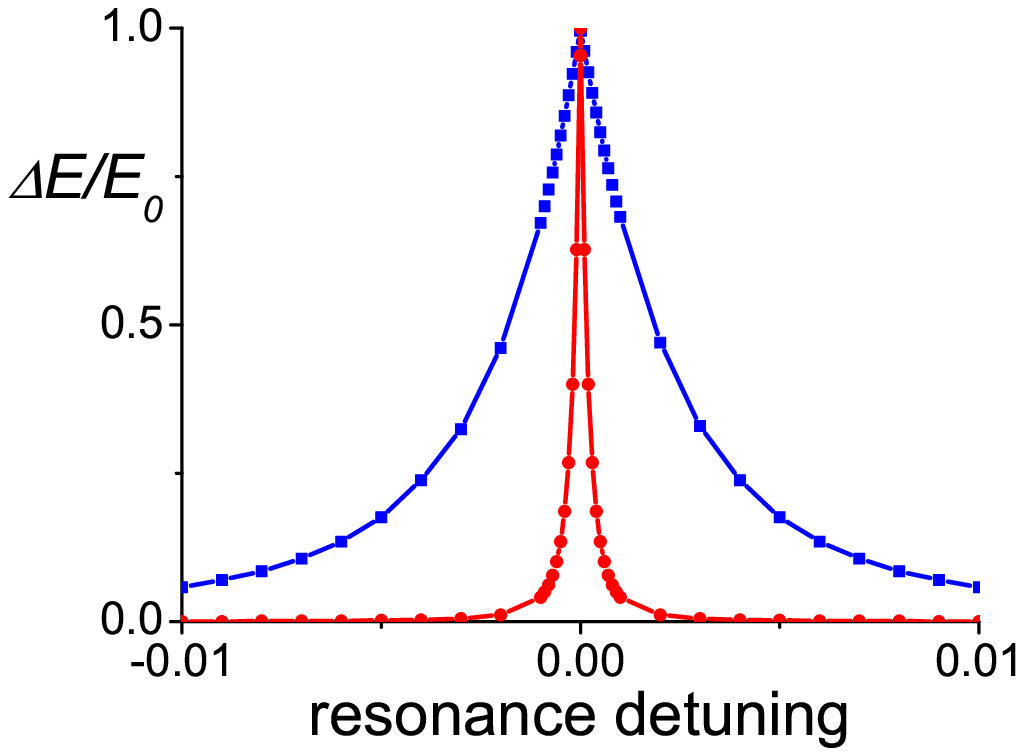}

\vspace{-0.6 cm}

a \hspace{6cm} b

\end{center}
 \vspace{-1.0 cm}
 \caption{ \label{fig_3}
(a) The dependence of the stable mode/oscillator life-times
$\tau_2$ on the detuning parameter values for the Morse lattice
and the oscillatory model in the vicinity of the exact resonance
point; (b) the energy exchange degree vs. the detuning parameter
for the Morse lattice and the oscillatory model. The unstable
mode/oscillator are initially excited with $E_0=10^{-6}$. Blue
color -- Morse lattice, red color -- oscillatory model in both
panels.}
\end{figure*}

The dependence of the stable mode/oscillator life-times vs.
values $\varepsilon_1$/$\mu_1$ is also the same for both models.
The dependence is logarithmic (see Fig.~\ref{fig_3}a), as
predicted by \eqref{Tf}. We also found that the unstable
oscillator life-time $\tau_1$ with very good accuracy varies as
$\tau_1 \sim 1/\lambda$ in a wide range $(10^{-2} \leq \lambda
\leq 10^2)$, where $\lambda$ is the coupling parameter  in
\eqref{ham-a}. The mode/oscillator life-times are defined as a
time intervals for full energy exchange. For example, $\tau_2$ is
a time interval between points A and B in Fig.~\ref{fig_2}d.
Similarly $\tau_1$ is a time interval between points B and C in
Fig.~\ref{fig_2}d.

Fig.~\ref{fig_3}b demonstrates the energy exchange degree $\Delta
E/E_0$ vs. $\varepsilon_1$/$\mu_1$ for both models in the middle
range of the resonance detuning parameters ($10^{-4} \lesssim
|\varepsilon_1| \lesssim 10^{-2}$ and $10^{-6} \lesssim |\mu_1|
\lesssim 10^{-3}$). One can see the extreme sensitivity of
life-times and energy exchange degree to the variations of
detuning parameters.

\begin{figure*}
\begin{center}
\includegraphics[height=60mm,angle=0]{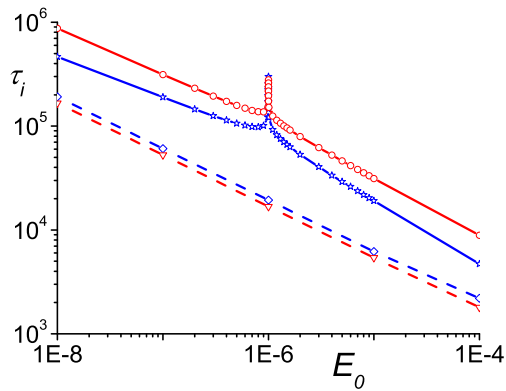}
\end{center}
\vspace{-1.0 cm} \caption{
 \label{fig_4} The dependence of modes/oscillators
life-times $\tau_i$ on the energy of the initially excited
unstable mode/oscillator $E_0$. Values
$\varepsilon_0=0{.}153\,808\,281\,3623$ and $\mu_0 =
3{.}374\,96152 \cdot 10^{-8}$ correspond to the exact resonance
point at $E_0 = 10^{-6}$ ($\varepsilon_1 = \mu_1 = 0{.}0$). Morse
lattice -- blue color, oscillatory model -- red color; solid
lines -- stable mode/oscillator, dashed lines -- unstable
mode/oscillator.}
\end{figure*}

As pointed above, the exact values for the onset of resonances
depend on the excitation energy $E_0$. Hence the energy exchange
kinetics should be also sensitive to $E_0$. Fig.~\ref{fig_4}
shows the dependencies of the life-times of both modes in the
Morse lattice and both oscillators vs. the energy of initially
excited unstable mode/oscillator $E_0$. The extreme sensitivity
of stable mode/oscillatory life-times in the vicinity of exact
resonance ($\varepsilon_0=0{.}153\,808\,281\,3623$ and $\mu_0 =
3{.}374\,961\,52 \cdot 10^{-8}$ at $E_0=10^{-6}$) is observed.

Thus, in the case of the initial excitation of unstable
mode/oscillator the resonance behavior is the same in both models
and is perfectly described by the analytical results in
Sec.~\ref{subsec:two-osc}. But the models have different features
in the case of stable mode/oscillator initial excitation. In the
oscillatory model the initially excited stable oscillator lives
infinitely long independently on the closeness to the resonance
point. The stable mode in the Morse lattice has finite life-time
and the energy exchange is observed, the resonance behavior being
independent on the resonance detuning. This different features of
the models is caused by the addition terms in the Morse potential
expansion which influence the energy exchange. Actually, an
addition of the term ${\widetilde \lambda} \, q_1 \, q_2^2$ in
\eqref{ham-a} results in the mode energy exchange and the finite
life-time of the stable vibrational mode.

\begin{figure*}
\begin{center}
\includegraphics[height=70mm,angle=0]{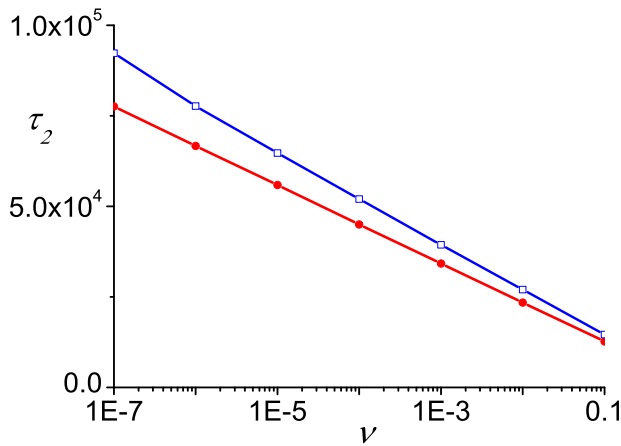}
\end{center}
\vspace{-1.5 cm} \caption{\label{fig_5} %
The dependence of the stable mode/oscillator life-times $\tau_2$
on the contribution of the unstable mode/oscillator energy $\nu$
into the total initial energy $E_0=10^{-6}$. The initial energies
of the stable and unstable modes are $(1-\nu)E_0$ and $\nu E_0$,
correspondingly; $\varepsilon_1 = \mu_1 = 0{.}0$. Blue color --
$\tau_2$ for the Morse lattice, red color -- $\tau_2$ for the
oscillatory model.
 }
\end{figure*}

The case of ``mixed'' initial conditions when the stable
mode/oscillator is excited with a small addition of the unstable
mode/oscillator is also analyzed and results are in good agreement
with analytical predictions (see \eqref{Tf}\,). Fig.~\ref{fig_5}
shows the dependence of the stable mode/oscillator life-times vs.
the contribution of the unstable mode/oscillator energy into the
total initial energy $E_0$. Here the initial energies of the
stable and unstable modes/oscillators are $(1-\nu)E_0$ and $\nu
E_0$, correspondingly.

Note that the observed resonance interaction of the vibrational
modes is the general phenomenon in random lattices under the
deformations at low temperatures. Actually an expansions of any
interaction potentials have the form ${\widetilde U} \sim x^2 \pm
\gamma \, x^3$. The minus sign means that the potential softens
under elongation (as in the Morse lattice). On contrary the plus
sign (as in the $\alpha$-FPUT lattice) means that the potential
softens under {\emph{contraction}}. This is the reason which
allowed to observe the similar resonance behavior in the random
FPUT lattice under contraction.


\section{Conclusions.}
 \label{sec:Concl}

In conclusion, we briefly summarize the main results.

\vspace{0.2 cm}

\noindent 1.  The vibrational modes resonance was found in the
random Morse lattice. Two features allow to observe the
resonances: the lattice randomness (random values of the
interaction potential; the random masses also can be used), and
the lattice deformation. At some values of the specific
elongation the exact resonance condition $\sum m_i \, \omega_i =
0$ are be fulfilled. The resonance $2 \omega_1 = \omega_2$ of two
most long-wave vibrational modes was chosen for the detailed
analysis. Small number of particles ($N=10$) and low excitation
energy allowed to avoid the resonances overlap.

\vspace{0.2 cm}

\noindent 2. Simple two-oscillators model with inharmonic
coupling was also considered. The analytical description of the
kinetics and modes energy exchange vs. the model parameters
(resonance detuning, specific energy) was performed.

\vspace{0.2 cm}

\noindent 3. Surprisingly good agreement was observed for the
kinetics and energy exchange rate for the Morse lattice and
simple model of two nonlinear coupled harmonic oscillators.

\vspace{0.2 cm}

\noindent 4. Modes/oscillatory life-times and the energy exchange
degree are extremely sensitive to the resonance detuning
parameters.

\vspace{0.2 cm}

\noindent 5. The resonance behavior is a general feature of
random nonlinear lattices at deformations and it was also
observed in the random $\alpha$-FPUT lattice.

\vspace{0.2 cm}

The observed resonance phenomenon can not throw some light on the
problem of energy equipartition and onset of the chaos because of
small values of $N$ and low specific energy, but it can serve as
a starting point for the analysis of initial stages of
vibrational modes interaction.


\vspace{0.5 cm}

This work was supported by the RFBR Grant \# 08--02--00253.

\end{document}